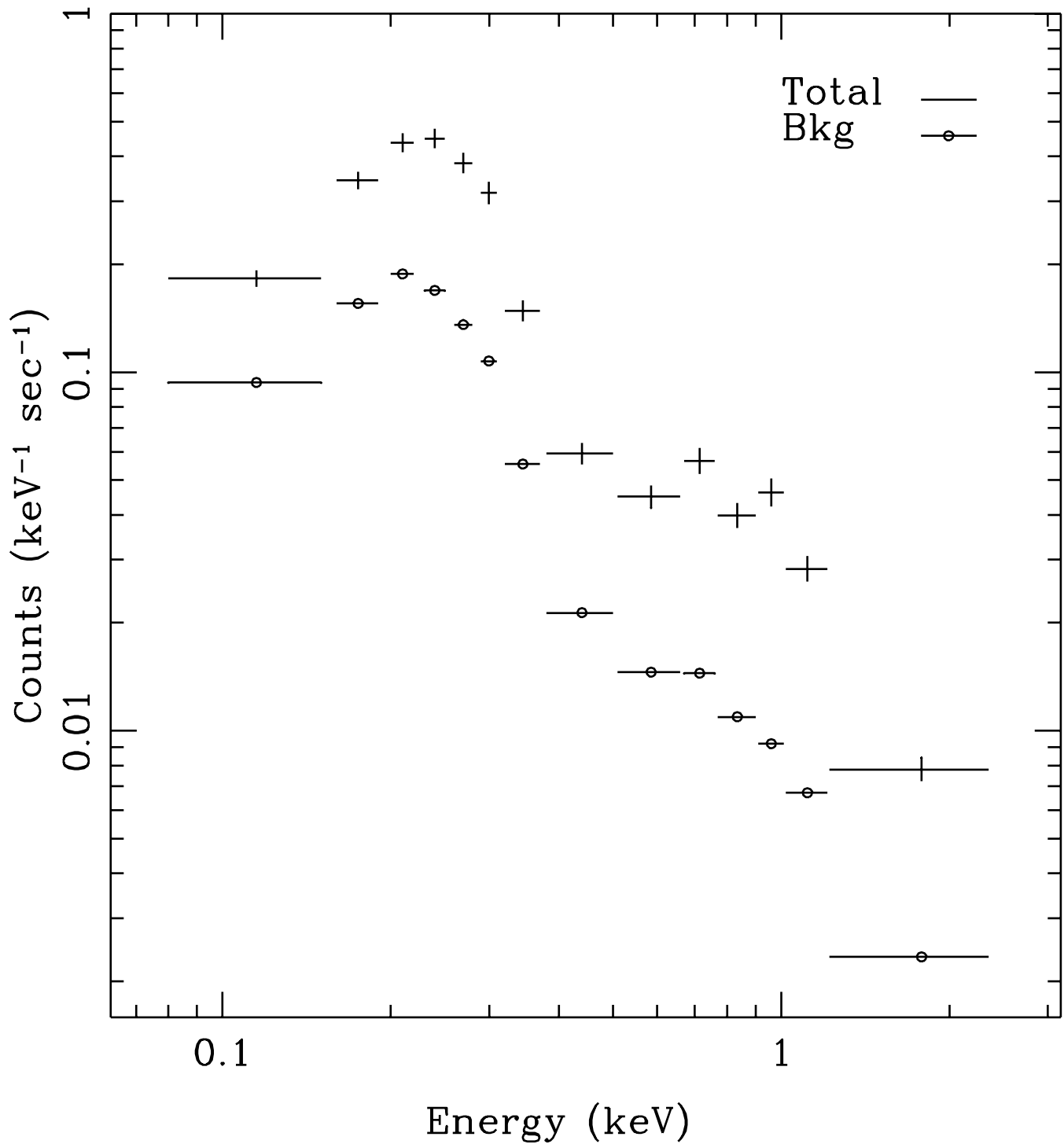

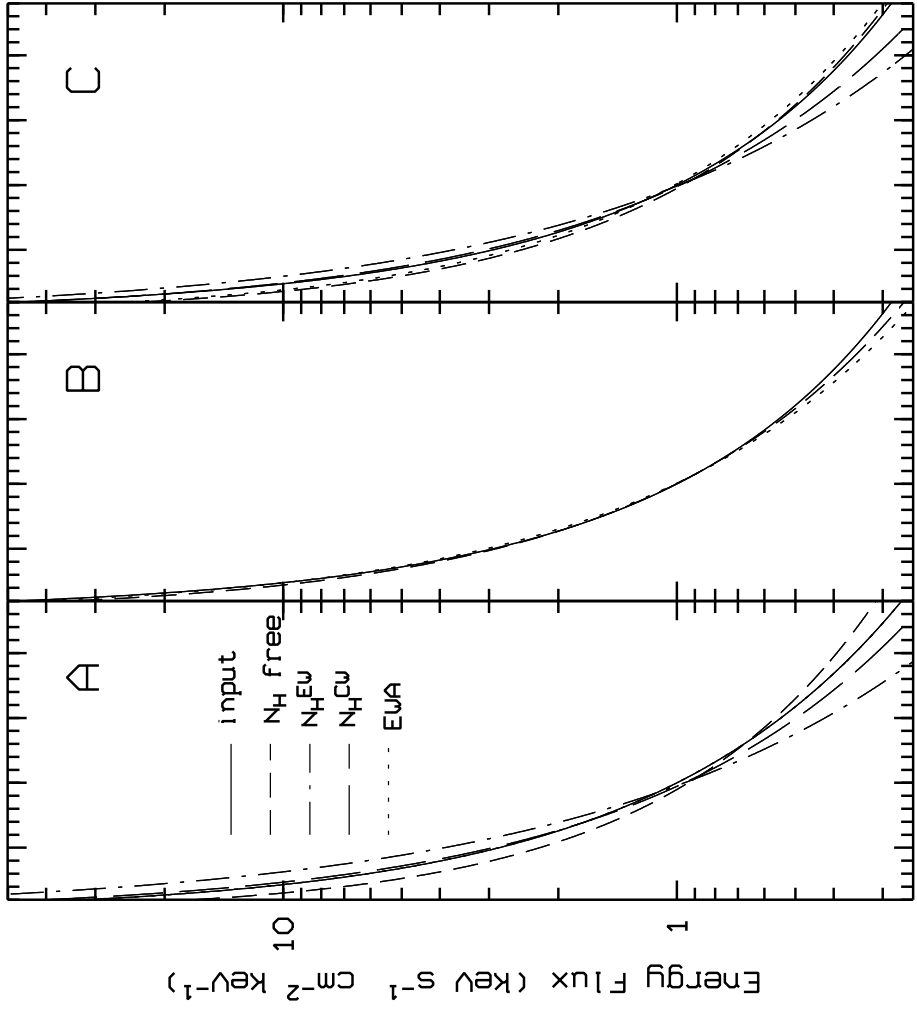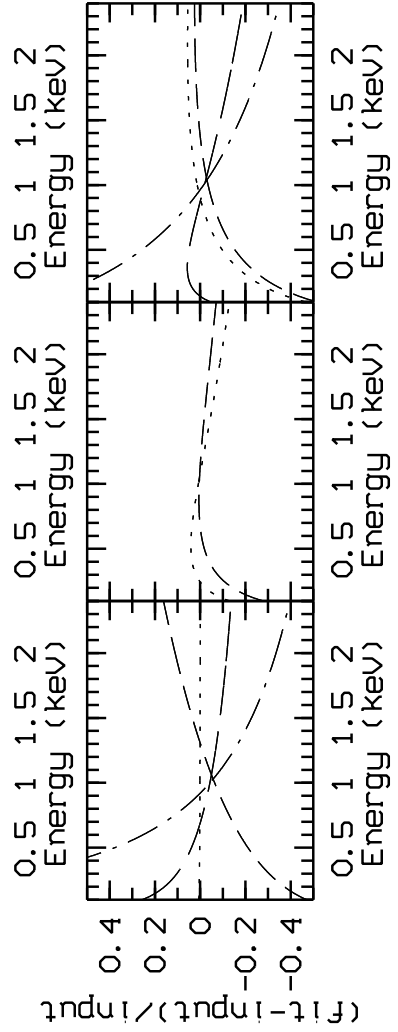

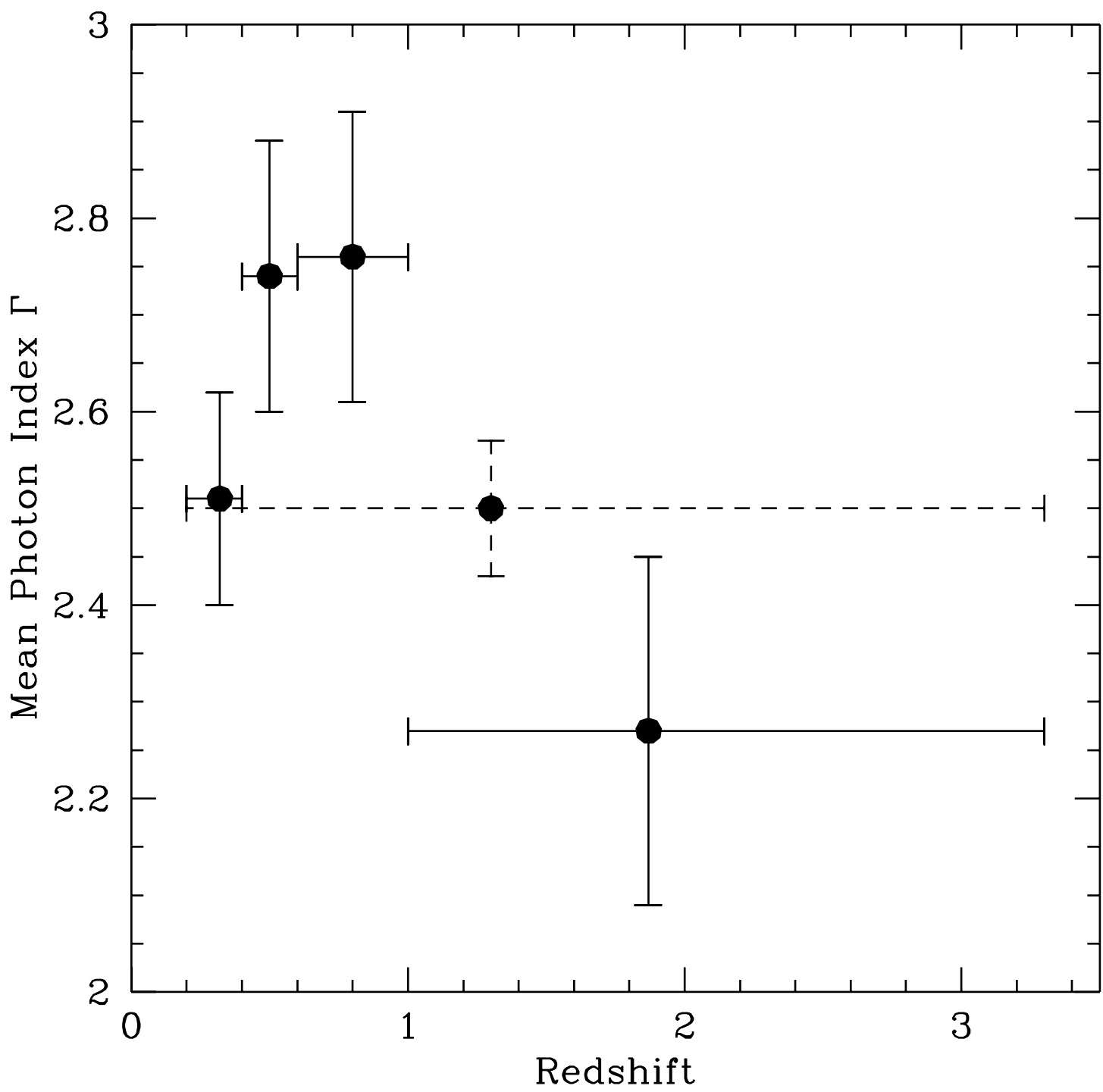

# *ROSAT* soft X-Ray properties of the Large Bright Quasar Survey: modeling of stacked X-Ray spectra

Norbert Schartel[1,2], Paul J. Green[3], Scott F. Anderson[4], Paul C. Hewett[5],

Craig B. Foltz[6], Bruce Margon[4], Wolfgang Brinkmann[2], Henner Fink[2],

and Joachim Trümper[2]

[1] *ESA, IUE Observatory, P.O. Box 50727, E-28080 Madrid, Spain. Affiliated to the Astrophysics Division, Space Science Department, ESTEC*

[2] *Max Planck Institut für extraterrestrische Physik, Giessenbachstr. 1, D-85740 Garching bei München, Germany*

[3] *Harvard-Smithsonian Center for Astrophysics, 60 Garden St., Cambridge, MA 02138*

[4] *Astronomy Department, University of Washington, Seattle, WA 98195*

[5] *Institute of Astronomy, University of Cambridge, Madingley Road, Cambridge CB3 0HA, UK*

[6] *Multiple Mirror Telescope Observatory, University of Arizona, Tucson, AZ 85721*

**ABSTRACT**

We develop and apply a novel method of analysis to study the X-ray spectral properties of 908 QSOs in the Large Bright Quasar Survey (LBQS) that were observed during the soft X-ray *ROSAT* All-Sky Survey (RASS). Due to the relatively short ($\leq 600\,\mathrm{s}$) RASS exposure times, only 10% of the QSOs are detected in X-rays, so X-ray spectral model fits for individual QSOs are precluded by poor photon statistics. Spectral stacking provides effectively much more sensitive X-ray observations for an *average* QSO in bins of redshift, and for several classes of QSOs. We model the stacked X-ray spectra in a way that obviates both the distinction between detections and non-detections, *and* the need to stack together only those objects observed through similar Galactic column densities $N_H$. In application to the LBQS/RASS sample, we marginally confirm a flattening of the X-ray spectral index of QSOs toward higher redshifts. Radio-loud QSOs show flatter X-ray photon indices, in agreement with previous work. We show that the apparent flattening of the photon index with redshift is not due to an increasing fractional contribution from the radio-loud class of QSOs.

# 1 INTRODUCTION

Although quasi-stellar objects (hereafter, QSOs) radiate similar amounts of energy in the soft X-ray and optical bandpasses, the soft X-ray properties of these objects are by comparison poorly understood. Since the vast majority of soft X-ray observations of QSOs are broadband measurements of relatively low spectral resolution, the overall shape of the continuum is critical to calculation of X-ray flux and luminosity through counts-to-flux conversion factors and the cosmological $k$-correction. These data are necessary for calculation of the intrinsic emission properties of QSOs, their X-ray luminosity function and its evolution with look-back time, and the contribution of QSOs to the diffuse X-ray background (XRB). The X-ray spectral slopes, and their connection to spectral slopes in other wavebands help to constrain the diverse physical models proposed to explain the continuum and line emission from the active galactic nucleus.

The comparatively low energy resolution of most X-ray satellites has limited the analysis of QSO X-ray spectra to simple models (usually power law plus absorption). Low photon flux has further limited these studies mostly to apparently bright, low-redshift QSOs. As a result, the QSOs with well-studied X-ray spectra have necessarily formed a highly incomplete sample, biased by a variety of poorly-understood selection effects. Previous efforts to minimize selection effects have concentrated on complete samples of low-redshift QSOs observed in targeted exposures or all-sky surveys (e.g., Laor et al. 1994, Walter & Fink 1993, respectively), or on deep pointings toward high redshift QSOs (e.g., Wilkes et al. 1992, Elvis et al. 1994a).

To ease the deconvolution of selection effects from intrinsic properties, and to enable robust statistical analysis, a large sample of QSOs selected using uniform, quantifiable criteria and with detailed multiwavelength information available is desirable. The Large Bright Quasar Survey (Hewett, Foltz & Chaffee 1994, Morris et al. 1991, Chaffee et al. 1991, Hewett et al. 1991, Foltz et al. 1989, Foltz et al. 1987) is a large, uniformly-selected sample of QSOs covering a wide range of redshifts. LBQS QSO candidates were selected from photographic and objective prism plates, using a combination of quantifiable selection techniques sensitive to colours, to strong emission lines, redshifted absorption features or continuum breaks. As described in the series of LBQS papers, the technique appears to be highly efficient at finding QSOs with $0.2 < z < 3.3$, a significantly broader range than past work. The entire LBQS optical sample consists of 1056 QSOs with $16.0 \leq B_J \leq 18.9$ over more than 450 deg$^2$ of the sky. Measurement of the ensemble X-ray properties of QSOs to redshifts greater than a few tenths has so far been thwarted by the cost and difficulty of making numerous deep targeted X-ray exposures. Two approaches are available to circumvent these problems using a large, homogeneous sample such as the LBQS.

One approach to determine the soft X-ray properties of such a sample is to search for deep X-ray images that serendipitously include sample QSOs. A random subsample of 146 LBQS QSOs included in *Einstein* IPC pointings was studied by Margon et al. (1992). The small number of counts for most of the QSOs in the LBQS/IPC overlap sample, along with uncertainties in the response corrections for off-axis sources, preclude the derivation of detailed spectral constraints for individual QSOs. They co-added X-ray images of all QSOs in similar ranges of Galactic hydrogen column density $N_H^{Gal}$ maintaining all the spectral and background information. The properties of the 'stacked' QSO are then analysed as if it were a single object, but measurements instead provide average values characterizing the subsample of co-added QSOs.

In a second approach, we study a much larger fraction of the LBQS sample now available for study using the first all-sky soft X-ray survey. The Position Sensitive Proportional Counter (PSPC, Pfefferman et al. 1987) onboard the *ROSAT* X-ray satellite (Trümper 1983, Aschenbach 1988) provides soft X-ray information for several thousand previously cataloged QSOs in the *ROSAT* All-Sky Survey (hereafter, RASS). The X-ray and optical properties of an LBQS/RASS overlap sample of 908 QSOs have been described and analysed by Green et al. (1995). Due to the short ($\leq 600$ s) RASS exposure times, only 10% of the QSOs are individually detected in X-rays. However, by stacking X-ray counts, they obtain effectively much more sensitive observations for an *average* QSO in bins of redshift or luminosity, and for several classes of QSOs. In the current work, we use stacking to study the soft X-ray *spectral* properties of several subsamples of the LBQS/RASS QSOs.

Since the RASS exposure times for LBQS QSOs are short, even stacking in bins of $N_H^{Gal}$ often does not yield a sufficient number of counts to allow X-ray spectral analysis. Consequently, we have developed and introduce here a novel technique that permits us to co-add QSO X-ray spectral data, whether detected or not, and even when observed through substantially different Galactic column densities. In summary, we first stack together X-ray counts preserving the pulse height channel (energy) information of each photon. The fit is performed using a multiplicative model that incorporates the absorption due to the Galactic column for each QSO, weighted according to its exposure time. This exposure-weighted absorption (hereafter, EWA) model is used throughout our analysis. To find the best-fit slope for a given spectral model, we fold the model through the PSPC response function for comparison to the data. If, after stacking, the signal-to-noise (S/N) ratio is still too low for a full least-squares fit to the observed X-ray spectrum, we use an adaptive hardness ratio (AHR) technique (see appendix § A) to derive the corresponding X-ray photon index. Our method is detailed in the following section (§ 2). We test the method with simulated high S/N data in § 3, and with low S/N data from the LBQS/RASS in § 4.1 Finally, we apply the EWA method to study the X-ray spectral properties of the LBQS QSOs in § 5 Readers most interested in the scientific results are directed to that section and the discussion following (§ 6).

## 2 X-RAY SPECTRAL ANALYSIS

### 2.1 The Spectral Stacking Method

To generate the stacked spectrum from a sample of $N$ sources, for simplicity we first describe the spectrum of a single source. As detailed in Green (1995), we used a source aperture radius of $180''$, optimized for optimal signal-to-noise of faint point sources in the RASS. A background annulus from 600 to $1200''$ is taken at each optical QSO position, from which unrelated detected sources are excised. In an exposure time $t_n^s$, the number of counts $C_{in}^q$ received from the $n$th source in the $i$th pulse height channel is given by:

$$C_{in}^q = \left( \sum_{k=1}^{K_n} V_{ink}^s - \frac{A_n^s \, t_n^s}{A_n^b \, t_n^b} \sum_{l=1}^{L_n} V_{inl}^b \right) \quad,$$

where these variables represent for the source [background] of the $n$th source, the aperture area or observed solid angle $A_n^s$ [$A_n^b$], the total number of photons $K_n$ [$L_n$], the vignetting correction factors $V_{ink}^s$ [$V_{inl}^b$] for the $k$th [$l$th] photon, and the exposure time $t_n^s$ [$t_n^b$]. The indices $q$, $b$ and $s$ pertain to the QSO, the background aperture, and the source aperture (total QSO plus background counts), respectively. PSPC background counts due to particles are not subject to vignetting as are X-rays, but are correctly subtracted here for point sources. This is especially justified in the case of stacked RASS data, because in the RASS scanning mode, the source and the background are not observed at different off-axis angles, resulting in the same mean vignetting correction.

The uncertainty in the net number of source counts $C_{in}^q$ is

$$\Delta C_{in}^q = \left( \sum_{k=1}^{K_n} (V_{ink}^s)^2 + \left( \frac{A_n^s \, t_n^s}{A_n^b \, t_n^b} \right)^2 \sum_{l=1}^{L_n} \left( V_{inl}^b \right)^2 \right)^{\frac{1}{2}} \quad .$$

The total exposure time $T^q$ is given by

$$T^q = \sum_{n=1}^{N} t_n^s \quad ,$$

where $N$ is the number of sources. By summing over all $N$ sources the counts in pulse height channels 8 - 240 (corresponding approximately to energies 0.08 to 2.4 keV) and dividing by the total exposure time $T^q$, we get the stacked source spectrum, $R_i^q = \{R_8^s, R_9^s, ...R_{240}^s\}$. The mean count rates for the individual channels for all $N$ sources in the sample is thus

$$R_i^q = \frac{1}{T^q} \sum_{n=1}^{N} \left( \sum_{k=1}^{K_n} V_{ink}^s - \frac{A_n^s \, t_n^s}{A_n^b \, t_n^b} \sum_{l=1}^{L_n} V_{inl}^b \right) \quad ,$$

with error

$$\Delta R_i^q = \frac{1}{T^s} \left( \sum_{n=1}^{N} \left( \sum_{k=1}^{K_n} (V_{ink}^s)^2 + \sum_{l=1}^{L_n} \left( \frac{A_n^s \, t_n^s}{A_n^b \, t_n^b} \right)^2 \left( V_{inl}^b \right)^2 \right) \right)^{\frac{1}{2}} \quad .$$

As an example, the stacked spectrum of all 72 detected QSOs in the LBQS/RASS sample with $0.2 < z < 1.4$ is shown in Figure 1.

### 2.2   Exposure-Weighted Absorption (EWA) Models

Because the absorption of soft X-rays by material along the line of sight to a QSO is energy-dependent and a non-linear function of the intervening column density, mean spectral parameters from stacking have been determined either by stacking together only QSOs with similar $N_H^{Gal}$ (e.g., Margon et al. 1992) using 21cm measurements (e.g. Stark et al. 1992), or by stacking $\chi^2$ contours from fits to individual QSOs (Canizares & White 1989). Here we outline a new method to derive the mean spectral parameters for an ensemble of sources where the individual sources need not be detected, and may be observed through a range of absorbing columns. Although the method we outline is suitable for a variety of X-ray spectral models (e.g., Raymond-Smith, thermal bremsstrahlung, blackbody) we assume here a power law index, which generally provides the best fit to the X-ray spectra of QSOs. With this assumption, the model spectrum of the $n$-th source is given by:

$$f_n\left(E, (N_H)_n, I_n, \langle\Gamma\rangle, E_0\right) \; dE = I_n \left(\frac{E}{E_0}\right)^{\langle\Gamma\rangle} e^{-(N_H)_n \sigma(E)} \; dE \quad, \tag{1}$$

where $f_n$ is the photon flux density of the $n$-th model spectrum in photons cm$^{-2}$ s$^{-1}$ keV$^{-1}$, $E_0$ is the reference energy in keV, $I_n$ is the model normalization (the photon flux density at $E_0$ in photons cm$^{-2}$ s$^{-1}$ keV$^{-1}$), $\langle\Gamma\rangle$ is the mean photon index, $(N_H)_n$ is the equivalent hydrogen column density of the $n$-th model spectrum (in cm$^{-2}$), and $\sigma(E)$ is the photoelectric cross-section following Morrison & McCammon (1983). We stress that the method is only rigorously correct if each input spectrum has a photon index equal to the mean $\langle\Gamma\rangle$. However, within the uncertainties, a mean index representative of the sample mean will result

The model for the stacked spectrum is given by the sum of all $N$ model spectra weighted by their exposure time:

$$F\left(E, (N_H)_n, I_n, \langle\Gamma\rangle, E_0, N\right) \; dE = \sum_{n=1}^{N} \left(\frac{t_n^s}{T^q} I_n \left(\frac{E}{E_0}\right)^{\langle\Gamma\rangle} e^{-(N_H)_n \sigma(E)}\right) \; dE \quad, \tag{2}$$

where $F$ is the photon flux density of the model for the stacked spectrum in photons cm$^{-2}$ s$^{-1}$ keV$^{-1}$. Using the definition of the mean of the normalization $\langle I \rangle$,

$$\langle I \rangle = \frac{1}{N} \sum_{n=1}^{N} I_n \quad,$$

one can write for $I_n$,

$$I_n = \langle I \rangle + \delta I_n \tag{3}$$

where

$$\sum_{n=1}^{N} \delta I_n = 0 \quad. \tag{4}$$

Substituting equation 3 into equation 2 gives

$$\begin{aligned} F\left(E, (N_H)_n, I_n, \langle \Gamma \rangle, E_0, N\right) \ dE &= \left( \sum_{n=1}^{N} \frac{t_n^s}{T^q} \langle I \rangle \left(\frac{E}{E_0}\right)^{\langle \Gamma \rangle} e^{-(N_H)_n \sigma(E)} \right) \ dE \\ &+ \left( \sum_{n=1}^{N} \frac{t_n^s}{T^q} \delta I_n \left(\frac{E}{E_0}\right)^{\langle \Gamma \rangle} e^{-(N_H)_n \sigma(E)} \right) \ dE \\ &= \langle I \rangle \left(\frac{E}{E_0}\right)^{\langle \Gamma \rangle} \left( \sum_{n=1}^{N} \frac{t_n^s}{T^q} e^{-(N_H)_n \sigma(E)} \right) \ dE \\ &+ \frac{1}{T^q} \left(\frac{E}{E_0}\right)^{\langle \Gamma \rangle} \left( \sum_{n=1}^{N} t_n^s \ \delta I_n \ e^{-(N_H)_n \sigma(E)} \right) \ dE \ . \end{aligned}$$

If $\delta I_n$ is independent of $(N_H)_n$ and $t_n^s$, then, using equation 4,

$$\left( \sum_{n=1}^{N} t_n^s \ \delta I_n \ e^{-(N_H)_n \sigma(E)} \right) \approx 0 \tag{5}$$

one gets the model of the stacked spectrum:

$$F\left(E, (N_H)_n, I_n, \langle \Gamma \rangle, E_0, N\right) \ dE \approx \langle I \rangle \left(\frac{E}{E_0}\right)^{\langle \Gamma \rangle} \left( \sum_{n=1}^{N} \frac{t_n^s}{T^s} e^{-(N_H)_n \sigma(E)} \right) \ dE \ . \tag{6}$$

The model of the stacked spectrum is therefore given by the convolution of the exposure-weighted absorption with a power law whose normalization is the mean of the individual source normalizations.

We note that to meet the requirement for EWA models that $I_n$ be independent of $N_H^{Gal}$ and $t_n^s$ (see equation 5), samples which are X-ray-selected must be treated with caution. For example, a sample consisting of sources required to have more than some number of X-ray counts will show a correlation between $I_n$ and $N_H^{Gal}$, since only those with larger intrinsic luminosities will meet the selection criterion. Such a problem may be partially alleviated by the randomizing factor of distance, but is best avoided by using samples selected in other wavebands, such as the LBQS.

## 3    TESTS WITH HIGH SIGNAL TO NOISE SIMULATED SPECTRA

### 3.1    Tested Models

Before turning to the relatively low S/N RASS observations of the LBQS QSOs, we first test our technique using high S/N simulated data. We simulate the *ROSAT* PSPC spectra of 4 individual QSOs assuming a variety of power law spectral slopes and intervening absorbing columns. The simulated spectra are then stacked as described above. Finally, we fit the stacked spectrum with 4 power law models to see how well each reproduces the original input spectral parameters. First, we test a power law with 3 free parameters (normalization $I$, $N_H$ and $\Gamma$, the 'free $N_H$' fit). Second, we use a model with absorbing column fixed at the exposure-weighted mean $N_H$. By this, we mean equation 1, but with $N_H$ replaced by $\langle N_H \rangle^{EW} = \sum_{n=1}^{N} \frac{t_n^s}{T^s}(N_H)_n$. Third, we test a model with the absorbing column fixed at the counts-weighted mean $N_H$. By this, we mean equation 1, but with $N_H$ replaced by $\langle N_H \rangle^{CW} = \sum_{n=1}^{N} \frac{C_n^q}{C^q}(N_H)_n$. The latter model can be used only when the true source counts are known for each QSO, i.e., when every object in the sample is detected. Finally, we test our own exposure-weighted absorption (EWA) model, described above, which can be used irrespective of sample detection fraction.

### 3.2 Simulated Samples

We are most often interested in investigating the X-ray spectra of samples with similar *non*-X-ray parameters. For example. we investigate below the dependence of the average QSO X-ray power law slope on radio-loudness and redshift. The investigation of spectral evolution with look-back time requires binning in redshift. (Such binning also prevents substantial degradation in spectral resolution from co-adding a wide range of restframe energies.) Since in any magnitude-limited sample, redshift and flux are strongly correlated, we infer that QSOs in a redshift bin will tend to have similar X-ray fluxes. Since it is generally known that $l_{opt}$ and $l_x$ are strongly correlated, this clustering in X-ray flux will obtain also for optically-selected sample such as the LBQS (Green et al. 1995). For the particular case of the RASS, the exposure times are also similar. For these reasons, we have chosen our primary simulated samples to reflect this by requiring fixed exposure time and intrinsic flux.

To test how well stacking and EWA models reproduce the simulated parameters in a variety of situations, we generate 3 representative samples of 4 QSOs, each with fixed intrinsic flux and exposure time $t$. Sample A ($t = 500$ s) shares a single power law slope ($\Gamma = -2.5$) but covers a range of $N_H$. Sample B QSOs ($t = 750$ s) have $N_H = 2.5 \times 10^{20}$ cm$^{-2}$ and a range of $\Gamma$. Sample C ($t = 750$ s) covers a range of both $\Gamma$ and $N_H$. We then fit the individual spectra. Due to the intentionally high S/N, the individual fits reproduce the model parameters to within < 1%. After fitting, the number of counts of each spectrum is adjusted so that each source has an intrinsic monochromatic flux $I$ at 1 keV of 1.0 photon cm$^{-2}$ s$^{-1}$ keV$^{-1}$. In Table 1, we show the model parameters and resulting counts for the 12 QSOs in the 3 samples.

Table 1. Input parameters for simulated QSO spectra.

| Sample A | | Sample B | | Sample C | | |
| ($\Gamma = -2.5$, vary $N_H$) | | ($N_H = 2.5^a$, vary $\Gamma$) | | (vary $N_H$, vary $\Gamma$) | | |
| $N_H^a$ | Counts | $\Gamma$ | Counts | $N_H^a$ | $\Gamma$ | Counts |
|---|---|---|---|---|---|---|
| 1.0 | 378693 | −2.00 | 218798 | 1.0 | −2.00 | 344151 |
| 3.0 | 174568 | −2.25 | 252554 | 7.0 | −2.25 | 138599 |
| 5.0 | 119193 | −2.75 | 370993 | 5.0 | −2.75 | 199497 |
| 7.0 | 96498  | −3.00 | 472399 | 3.0 | −3.00 | 390066 |

[a] Intervening column density in units $10^{20}$ cm$^{-2}$.

After stacking the spectra in each sample according to the prescription in § 2.1, we then fit each stacked spectrum using 4 power law models (with free $N_H$, with exposure- and counts-weighted $\langle N_H \rangle$, and with EWA). Figure 2. illustrates these fits, with the best fit parameters for each sample/model combination listed in Table 2.

For sample A, EWA clearly provides the best fit method. EWA models are thus particularly well-suited to modeling stacked spectra of sources expected to have similar intrinsic spectral properties, when observed at many different sky positions or through patchy absorption.

For sample B, EWA and $\langle N_H \rangle$ models yield identical results as expected, since the effective absorption in both models is identical. The model with $N_H$ free inaccurately reproduces the input $N_H$, but appears to reproduce the input $\Gamma$, and $I$ parameters somewhat more accurately. Sample B might be approximated by a sample of QSOs observed in a single deep X-ray image. In this case, we recommend a model with $N_H$ free to best measure $\Gamma$, but only if there are adequate counts for a two-parameter fit.

Table 2. Power law fits to stacked simulated samples.

| Sample | Absorption Model | $N_H$ ($10^{20}$ cm$^{-2}$) | $I$ (cm$^{-2}$ s$^{-1}$ keV$^{-1}$) | $\Gamma$ | $\chi^2$/d.o.f. |
|---|---|---|---|---|---|
| | | | Best Fit Parameter[a] | | |
| A | free $N_H$ | $1.84 \pm 0.02$ | $0.936 \pm 0.003$ | $-2.253 \pm 0.009$ | 55.5 / 25 |
| A | $\langle N_H \rangle^{EW}$ | 4.0 | $0.957 \pm 0.003$ | $-3.026 \pm 0.003$ | 800.2 / 26 |
| A | $\langle N_H \rangle^{CW}$ | 2.83 | $0.957 \pm 0.003$ | $-2.618 \pm 0.003$ | 242.0 / 26 |
| A | EWA | ... | $1.001 \pm 0.003$ | $-2.502 \pm 0.002$ | 18.5 / 26 |
| B | free $N_H$ | $2.29 \pm 0.02$ | $1.004 \pm 0.003$ | $-2.521 \pm 0.007$ | 7.1 / 25 |
| B | $\langle N_H \rangle^{EW}$ | 2.5 | $1.008 \pm 0.002$ | $-2.600 \pm 0.002$ | 22.0 / 26 |
| B | $\langle N_H \rangle^{CW}$ | 2.5 | $1.008 \pm 0.002$ | $-2.600 \pm 0.002$ | 22.0 / 26 |
| B | EWA | ... | $1.008 \pm 0.002$ | $-2.600 \pm 0.002$ | 22.0 / 26 |
| C | free $N_H$ | $2.43 \pm 0.02$ | $0.965 \pm 0.003$ | $-2.364 \pm 0.007$ | 12.8 / 25 |
| C | $\langle N_H \rangle^{EW}$ | 4.0 | $0.985 \pm 0.002$ | $-2.907 \pm 0.002$ | 545.0 / 26 |
| C | $\langle N_H \rangle^{CW}$ | 3.25 | $0.981 \pm 0.002$ | $-2.651 \pm 0.002$ | 172.1 / 26 |
| C | EWA | ... | $1.013 \pm 0.002$ | $-2.387 \pm 0.002$ | 112.8 / 26 |

[a] Errors are taken from the covariance matrix of the fit and correspond to the 68.3 % confidence level.

In the most general case, Sample C, where both objects and sight-lines may have a dispersion in properties, EWA does not provide the best fit as judged by $\chi^2$ values, but most accurately retrieves the true $\langle \Gamma \rangle$. If all sources are detected, the intrinsic properties of the sample may be adequately understood by simply assuming a count-weighted mean $N_H$. But the $\langle \Gamma \rangle^{CW}$ model cannot be used when some sources have only X-ray flux upper limits, as for the LBQS/RASS sample.

We have also tested an analogous set of samples, where individual QSO spectra may have different exposure times, but are generated to have the same total counts and flux normalization. For such samples, EWA again provides best-fit spectral parameters closest to the input simulated spectra.

Overall, the EWA model performs as well or better than the other 3 power-law models tested for these samples, and is clearly superior when considering sources with few counts. We expect that strong average spectral constraints for a variety of object classes (not just QSOs) might be achieved by stacking spectra of samples observed in the large number of *Einstein* or *ROSAT* deep *pointed* observations now becoming available in data archives, and by using EWA models to correctly account for intervening absorption.

### 3.3   Absorption: Intrinsic or Galactic?

For no QSO in the LBQS/RASS sample are there sufficient counts for adequate power law spectral fits to constrain both $\Gamma$ and the total line-of-sight column, $N_H^{tot}=N_H^{intr}+N_H^{Gal}$. Thus, we (and many other workers) proceed with the assumption that the intrinsic column is negligible in the aggregate. The majority of optically-selected QSOs observed in soft X-rays to date indeed show little evidence for strong absorbing columns $N_H^{intr}$ intrinsic to the QSO (e.g., Fiore et al. 1994, Walter & Fink 1993, Canizares & White 1989, Wilkes & Elvis 1987). We verify this for the LBQS/RASS sample in § 4.1, and hereafter perform all model fits with absorption fixed toward individual sources at the Galactic value (taken from Stark et al. 1992), leaving two free parameters, the photon index $\Gamma$, and the flux normalization $I$.

## 4   TESTS WITH LOW SIGNAL TO NOISE DATA

### 4.1   Detection status and absorption

Green et al. (1995) describe criteria for detection of LBQS QSOs (or other faint point sources) observed in the RASS, optimized for maximum S/N. When stacking, we co-add X-ray counts from all QSOs regardless of their detection status, so we should derive similar spectral indices when stacking together samples of either detected or non-detected QSOs, if the intrinsic spectral behavior does not depend strongly on the measured count rate. Furthermore, if the derived spectral parameters differ for subsamples of QSOs observed through different $N_H^{Gal}$, this would suggest that fitting with an EWA model does not correctly treat the effects of absorption.

As a check on our technique, we thus divide the LBQS/RASS sample into subsamples, one with $1.5 < N_H^{Gal} \leq 2.5$ (in units of $10^{20}$ cm$^{-2}$), a second with $2.5 < N_H^{Gal} \leq 3.5$, and a third containing QSOs regardless of $N_H^{Gal}$. We further divide these according to detection status before stacking. Table 3 shows for each of these subsamples the number of QSOs, the range and mean of the Galactic column density $N_H^{Gal}$, the stacked count rate (counts per 1000 s), and the mean photon index from the best-fit EWA model (see § 2.2). Since there may be an intrinsic evolution of the mean X-ray photon index $\langle \Gamma \rangle$ with redshift, and since higher redshift QSOs are much less likely to be detected, we include only QSOs with $0.2 \leq z \leq 1.0$. Figure 1. shows the spectrum that results from stacking all 72 detected QSOs in this redshift range. Because the total number of photons in the stacked spectra of non-detected QSOs is small, we determine the spectral parameters of these using the adaptive hardness ratio technique (see the appendix).

We find that the best-fit photon index is, within the uncertainties, independent of either the division into absorption groups or the classification by detection status. A comparison of photon indices and their errors in Table 3 shows that the EWA method yields consistent photon indices across bins of Galactic column density (e.g., for the classes 1, 2, 4, and 5). A comparison between the mean photon indices of the detections, $2.65 \pm 0.07$ (class 3), and the non-detections, $2.56 \pm 0.11$ (class 6), also shows no significant difference. As expected, our results are thus independent of count rate and intervening Galactic absorption. The EWA model can be rigorously applied to the stacked spectra of QSOs observed with a wide range of exposure times or through a wide range of Galactic column.

We also performed fits to the stacked spectra including the possibility of absorption intrinsic to the QSOs. To do this, we convolve a power law with intrinsic absorption with the exposure-weighted Galactic absorption. The relatively small number of counts yields poor statistics, but fits were consistent at the $1\sigma$ level for all of the above classes having no intrinsic equivalent hydrogen column density (i.e., $N_H^{tot} \approx N_H^{Gal}$), justifying our explicit assumption of this in following analysis.

Table 3. Detection status and Galactic absorption.

| Class No. | No. of QSOs[a] | $N_H^{Gal}$ [b] Range | Mean | Count rate (per ksec) | $\langle\Gamma\rangle$ [c] (EWA) | $\chi^2_{red.}$ / d.o.f. |
|---|---|---|---|---|---|---|
| | | Detections | | | | |
| 1 | 37 | 1.5 - 2.5 | 1.99 | 73.5 ± 2.1 | 2.68±0.10 | 1.74 / 12 |
| 2 | 26 | 2.5 - 3.5 | 3.03 | 59.3 ± 3.2 | 2.61±0.13 | 1.23 / 11 |
| 3 | 72 | all | 2.53 | 66.9 ± 2.0 | 2.65±0.07 | 1.06 / 15 |
| | | Non-detections[d] | | | | |
| 4 | 129 | 1.5 - 2.5 | 2.05 | 10.5 ± 1.1 | 2.62±0.16 | AHR |
| 5 | 89 | 2.5 - 3.5 | 3.05 | 9.4 ± 1.2 | 2.46±0.20 | AHR |
| 6 | 275 | all | 2.81 | 9.6 ± 0.7 | 2.56±0.11 | AHR |

[a] Only QSOs with $z < 1$ included in this analysis.

[b] Galactic column density in units $10^{20}\,\mathrm{cm}^{-2}$.

[c] Errors are taken from the covariance matrix of the fit and correspond to the 68.3 % confidence level.

[d] Photon indices for non-detections calculated using the adaptive hardness ratio (AHR) technique.

# 5   SPECTRAL CHARACTERISTICS OF THE LBQS/RASS SAMPLE

## 5.1   Mean photon indices for the LBQS: RASS vs. *Einstein*

Stacking the spectra of the entire LBQS/RASS sample of 908 QSOs achieves a single spectrum with more than 6000 counts. Were so many counts available for a single QSO, or for an ensemble of QSOs in a very narrow redshift range, somewhat more detailed spectral analysis would be possible. The stacked spectrum of a sample such as the LBQS encompasses a wide range of redshifts, blurring out many spectral details. Since we are investigating assumed power law slopes that by their nature are conserved with respect to redshift, our derived mean X-ray photon index $\langle\Gamma\rangle$ should nevertheless adequately represent that of the mean QSO in the LBQS. We thus perform a standard $\chi^2$ fit to the stacked spectrum, assuming no intrinsic absorption. The result (with its error) is $\langle\Gamma\rangle = 2.53 \pm 0.07$ (listed in the last row of Table 4). This is consistent with the mean (with its dispersion) of $\langle\Gamma\rangle = 2.5 \pm 0.5$ for a sample of 58 low redshift ($\langle z \rangle = 0.08$) AGN measured in the RASS by Walter & Fink (1993). An initial PSPC spectral survey of 10 PG quasars by Laor et al. (1994) yields $\langle\Gamma\rangle = 2.5 \pm 0.4$, also consistent with our much larger sample.

We may also compare our derived $\langle\Gamma\rangle$ for the entire LBQS/RASS with that of the serendipitous *Einstein* sample (the LBQS/IPC sample; Margon et al. 1992), since the latter includes a random subsample of the former. This is confirmed by the similarity of the redshift distributions for the LBQS/RASS sample (see Table 4) to that of the *Einstein* sample, where $0.2 < z < 3.1$, and $\langle z \rangle = 1.33$. The mean photon index of the LBQS/RASS sample, $\Gamma_{RASS} = 2.53 \pm 0.07$, is measured between 0.08 keV and 2.4 keV, while the LBQS/IPC analysis yields $\Gamma_{IPC} = 2.3 \pm 0.2$, measured between 0.16 keV – 3.5 keV (both bandpasses in the observer's frame). We thus find no significant difference between the mean photon index of the LBQS QSOs using *Einstein* and RASS spectra, but the combined uncertainty is large.

A variety of papers (e.g., Schartel, Walter, & Fink 1993, Brunner et al. 1992, Laor et al. 1994) have found steeper (softer) power law slopes from *ROSAT* PSPC than *Einstein* IPC data. Fiore et al. (1994) demonstrate that fits to IPC spectra of QSOs yield indices lower on average by $0.5 - 0.7$ compared to PSPC spectra *even for the same objects*. They argue that some of this discrepancy may be due to calibration errors in the response matrices of one or both of the IPC and PSPC, which may weaken the the significance of direct PSPC/IPC comparisons.

### 5.2 Redshift subsamples

We have derived the photon index for stacked subsamples of LBQS/RASS QSOs divided into 4 bins of redshift, with bin edges $z = 0.2, 0.4, 0.6, 1.0$, and 3.3. We chose these particular bins to optimize comparison with results for RQ and RL QSOs in § 5.3, but we note that our results are not significantly affected by the exact choice of redshift bin boundaries. The mean photon indices by redshift are shown in Table 4, along with the overall mean photon index from EWA models for the entire LBQS/RASS sample.

Table 4. Redshift.

| Class No. | No. of QSOs | Redshift Range | Redshift Mean | $T_{total}$ (ksec) | Count rate (per ksec) | $\langle \Gamma \rangle^a$ (EWA) | $\chi^2_{red.}$ / d.o.f. |
|---|---|---|---|---|---|---|---|
| 1 | 85  | 0.2 - 0.4 | 0.32 | 47.9  | 34.4 ± 1.5 | 2.51 ± 0.11 | 1.16 / 10 |
| 2 | 89  | 0.4 - 0.6 | 0.50 | 51.3  | 23.3 ± 1.4 | 2.74 ± 0.14 | 0.68 / 8 |
| 3 | 174 | 0.6 - 1.0 | 0.80 | 96.5  | 15.1 ± 0.9 | 2.78 ± 0.15 | 0.88 / 4 |
| 4 | 560 | 1.0 - 3.3 | 1.72 | 316.7 | 5.8 ± 0.5  | 2.28 ± 0.18 | 0.98 / 4 |
| 5 | 348 | 0.2 - 1.0 | 0.61 | 195.7 | 22.0 ± 0.7 | 2.66 ± 0.07 | AHR |
| 6 | 908 | 0.2 - 3.3 | 1.29 | 512.4 | 12.0 ± 0.4 | 2.53 ± 0.07 | AHR |

$^a$ Errors are taken from the covariance matrix of the fit and correspond to the 68.3 % confidence level.

Figure 3. shows the mean ROSAT photon index fits to stacked spectra from our four redshift subsamples (filled circles with solid bars). The width of the horizontal bars represents the redshift bin range for stacking. The mean photon index for the entire LBQS/RASS sample is also plotted (filled circle, dashed bars). The largest observed difference is a decrease at the $\approx 2\sigma$ level between the photon indices for classes 3 and 4, with mean redshifts $\langle z \rangle = 0.8$ and 1.9, respectively. To reduce the uncertainty in the mean low redshift index, we also tried a broader low-redshift subsample ($0.2 < z < 1.0$; class 5), for which we still find a difference of similar significance.

The decrease in measured photon index with redshift is similar to that of Stewart et al. (1994) from PSPC spectra in 5 deep ROSAT fields, containing about 100 UVX QSOs (Boyle et al. 1990). As Stewart et al. pointed out, the power-law component of QSOs to $z \lesssim 3$ does not attain the medium-energy slope of the cosmic XRB over the 2 - 10 keV range ($\Gamma = 1.4$;

see Holt 1992 for a review), even though the rest-frame bandpass is similar. Thus QSOs cannot form 100% of the XRB. Autocorrelation of the XRB also suggests that if the remainder is to come substantially from point sources, they must have higher surface densities than QSOs (Georgantopolous 1993). Narrow emission-line galaxies may provide both the harder spectra and higher surface densities required (e.g., Shanks et al. 1991, Jones et al. 1995).

From a large sample of serendipitous detected sources in archived *ROSAT* PSPC pointings, Vikhlinin et al. (1995) present a study of X-ray spectral slope as a function of source flux. They find that, after some corrections for Galactic stellar contamination, the mean photon index decreases monotonically toward lower fluxes. A direct extrapolation of that trend to fluxes of about $10^{-15}$ erg cm$^{-2}$ s$^{-1}$ is consistent with the slope of the cosmic XRB, and includes all point-source contributors. Since Vikhlinin et al. (1995) invoke X-ray selection, however, we suspect that (the more X-ray luminous) RL QSOs may make a larger fractional contribution to their derived mean photon index (and also to the XRB), which in turn would yield lower values of $\langle \Gamma \rangle$. Even though the LBQS/RASS sample is optically-selected, we confirm the hardening trend observed by Vikhlinin et al. in the flux decade where our samples overlap, between $10^{-12}$ and $10^{-13}$ erg cm$^{-2}$ s$^{-1}$. (The counts-to-flux conversion factor for our redshift subsamples in Table 4 is about $2.2 \times 10^{-11}$ cts$^{-1}$ erg cm$^{-2}$.) The large, deep sample of Vikhlinin et al. (1995) is complementary to the current study, and is clearly an important contribution to the ongoing debate as to the origin of the cosmic XRB. However, the importance of complete optical spectroscopic followup should not be underestimated.

### 5.3  The radio-loud and the radio-quiet subsample

Optically-selected samples of QSOs appear to be separated into two distinct populations, radio-loud (RL) and radio-quiet (RQ; e.g., Kellerman et al. 1989). Although these populations have quite similar optical properties, the X-ray emission from RL QSOs is now generally accepted to be stronger than that from RQ QSOs (e.g., Avni & Tananbaum 1986, Worrall et al. 1987). In addition, there appears to be a significant difference between the mean X-ray photon indices of heterogeneous samples of radio-loud and radio-quiet QSOs (e.g., Shastri et al. 1993, Molendi et al. 1992, Wilkes & Elvis 1987). We are now able to examine the *ROSAT* X-ray spectral properties of a substantial fraction of the LBQS sample, divided by radio-loudness.

Table 5. Radio loudness.

| Class | No. of QSOs | $\langle z \rangle$ | $T_{total}$ (ksec) | Count rate (per ksec) | $\langle \Gamma \rangle^a$ (EWA) | $\chi^2_{red.}$ / d.o.f. |
|---|---|---|---|---|---|---|
| Radio-loud | 40 | 1.22 | 22.5 | $21.5 \pm 2.1$ | $2.15 \pm 0.23$ | 0.41 / 3 |
| Radio-quiet | 147 | 1.38 | 85.4 | $13.3 \pm 1.0$ | $2.65 \pm 0.18$ | 1.17 / 5 |

$^a$ Errors are taken from the covariance matrix of the fit and correspond to the 68.3 % confidence level.

Radio data are currently available for 226 of the QSOs in the LBQS/RASS sample from VLA observations at 8.4 GHz (Visnovsky et al. 1992, Hooper et al. 1995), and from 26 (much less sensitive) detections from the sky surveys of Condon et al. (1988, 1991) at 5 GHz, or Large et al. (1981) at 408 MHz. RASS data for this latter radio-loud sample was recently analysed by Brinkmann, Siebert, & Boller (1994) and Brinkmann & Siebert (1995). We define an object as radio-loud if the log ratio of the emitted monochromatic luminosities at 8.4GHz and 4410 Å, $\log R \equiv \log (l_{8.4\,\mathrm{GHz}}/l_{4410\mathrm{\AA}})$, is greater than unity, as described in Green et al. (1995). This definition yields 193 radio-quiet and 42 radio-loud QSOs. There is only 1 RL QSO below $z = 0.4$, compared with 20 RQ QSOs. Furthermore, $z = 0.4$ corresponds to the mean redshift of QSOs with $M_B = -24$: Hooper et al. (1995) and Della Ceca et al. (1994) find that the fraction of RL QSOs decreases abruptly for absolute magnitudes fainter than this. There is also only 1 RL QSO above $z = 2.4$, compared to 26 RQ QSOs. For these reasons, to enable a fair comparison between the RL and RQ QSO samples, we restrict the redshift range of both samples to $0.4 < z < 2.4$. This leaves 40 RL QSOs (with a mean and r.m.s. for the redshift distribution of $\langle z \rangle = 1.2 \pm 0.5$), and 147 RQ QSOs ($\langle z \rangle = 1.4 \pm 0.6$).

Table 5 briefly characterizes the RL and RQ QSO samples and power law model fits to their stacked spectra as follows: the mean redshift, the total exposure time, number of counts per 1000 s, and the photon index as derived assuming in the EWA models that $N_H^{tot} = N_H^{Gal}$. The RQ QSOs show a steeper mean spectrum ($\langle \Gamma \rangle = 2.7 \pm 0.2$) than the RL QSOs ($2.2 \pm 0.2$), but only at the $1.7\sigma$ level. Although the RL QSOs in the LBQS/RASS show higher count rates (at $3.5\sigma$, see Table 5), since RL QSOs are fewer in number, their total number of counts is smaller by a factor of about 2.5, resulting in larger uncertainties in their mean photon index $\langle \Gamma \rangle$. Our results for RL and RQ QSO samples are consistent with the *ROSAT* PSPC photon indices found by other authors when similar redshift ranges are covered. These include studies from smaller and more heterogeneous samples of pointed observations (e.g., Fiore et al. 1994, Laor et al. 1994), as well as studies using RASS detections (e.g., Brunner et al. 1992, Schartel, Walter, & Fink 1993, Brinkmann, Siebert & Boller 1994). Unlike the current work, all these studies suffer from X-ray selection effects, but they have the advantage of better overall photon statistics.

### 5.4 QSOs with strong FeII emission

Emission lines in QSOs are attributed to photoionization of gas by the high energy continuum near the QSO nucleus. The details of the process are not well-understood, but may be illuminated by comparison of emission line to continuum properties in samples of QSOs. Based on a heterogeneous sample of 18 objects with $\langle z \rangle = 0.15$, Shastri et al. (1993) found that (radio-quiet) QSOs with strong FeII $\lambda 4570$ emission show softer (steeper) *Einstein* spectral slopes. We may contrast this result with X-ray spectral constraints on a subsample of the LBQS/RASS QSOs showing strong UV FeII emission. We use the subsample defined in Green et al. (1995), consisting of 23 QSOs in the LBQS/RASS sample flagged as having strong UV FeII emission. Although the FeII flag represents a subjective judgment, these QSOs are likely to be among those with the strongest iron emission in the LBQS sample. The redshift range most likely to lead to their FeII classification is $0.4 \leq z \leq 1.5$, due to the iron feature under [Ne IV] $\lambda 2423$. We thus restrict the redshifts of both the FeII and a comparison non-FeII sample to within this range. The mean redshift of the FeII sample is $\langle z \rangle = 0.88$. The comparison sample of 467 non-FeII QSOs has a similar mean redshift of 0.96.

Using an EWA model and the AHR technique, we derive for FeII-strong QSOs a mean photon index $\langle \Gamma \rangle = 2.8 \pm 0.2$, consistent within the uncertainties with our result for the non-FeII sample, $\langle \Gamma \rangle = 2.7 \pm 0.1$.

Our sample is probably more uniformly-selected that that of Shastri et al. (1993). Our sample has a wider range of redshifts, with a much larger mean redshift, and our classification is based on rest-frame UV, not optical, iron lines. The sample of Shastri et al. is limited to RQ QSOs only. Twelve of our 23 FeII-strong QSOs have radio data, of which 1148–0033 and 0107-0235 are radio-loud QSOs; only the latter is detected in the RASS. The exclusion of these 2 quasars from the stacked spectra does not significantly change our results. Neither would the exclusion of one or two more RL QSOs possibly included in the 11 QSOs without radio data, since no single QSO dominates the counts in the stacked spectrum. Since different energy ranges and different iron lines are sampled, we consider our null result to complement rather than contradict the results of Shastri et al. More detailed analysis, including correlations of FeII equivalent width and X-ray luminosity, are pending.

# 6 DISCUSSION

What might cause the apparent flattening (hardening) of the 0.08 keV – 2.4 keV X-ray slope we find toward higher redshifts? We consider briefly several possible explanations of the observed decrease in $\langle\Gamma\rangle$ that occurs somewhere between redshift 1 and 2. We order this discussion by considering effects increasingly linked to the intrinsic physics of the QSO nuclear continuum.

1) An increasing contribution from radio-loud QSOs. Since RL QSOs are known to have flatter X-ray spectra, and to be generally more luminous in X-rays than their RQ counterparts, the observed flattening of $\langle\Gamma\rangle$ toward higher redshift might simply be due to an increased contribution of X-ray counts from RL QSOs at higher redshift. Since stacking eliminates the distinction between detections and non-detections, this trend would *not* be caused by the larger detection fraction of the more X-ray-luminous RL QSOs. However, there is some evidence (e.g., Hooper et al. 1995) that RL QSOs may represent an increasing fraction of all QSOs at high redshift. Also, the study of Green et al. (1995) suggested for the LBQS/RASS sample that RQ QSOs show a relative decrease in X-ray luminosity with redshift not seen in RL QSOs. The fractional X-ray contribution of RL QSOs thus increases, possibly leading to the observed decrease of $\langle\Gamma\rangle$ with redshift. We test this possibility in two ways below.

First, we examine whether trends in $\langle\Gamma\rangle$ exist within the RL and RQ populations separately. For a sample of RASS-detected RL QSOs, Schartel, Walter, & Fink (1993) indeed found a decrease in $\langle\Gamma\rangle$. In their highest redshift bin, ($\langle z\rangle\approx 1.7$) they find $\langle\Gamma\rangle = 1.5 \pm 0.2$. It is intriguing that this value approaches that of the low-flux sources in Vikhlinin et al. (1995), and that of the XRB as well. Do RQ QSOs alone show a trend in $\langle\Gamma\rangle$ with redshift? Since the LBQS/RASS sample of RQ QSOs is relatively large, we may stack their RASS spectra in 3 redshift bins, with edges at $z = 0.2, 0.6, 1.0, 3.3$. The photon indices for these 3 bins (with $\langle z \rangle = 0.40$, 0.77, and 1.87) derived using the AHR technique are $2.7 \pm 0.1$, $3.2 \pm 0.4$, and $2.4 \pm 0.3$, respectively. Although there is some suggestion that the hardening occurs even for RQ QSOs alone, the apparent decrease in $\langle \Gamma \rangle$ is only at about the $1.4\sigma$ level.

Second, we estimate the fraction of X-ray photons in each redshift bin that are from RL QSOs. Let $R_{RL} = N_{RL}/(N_{RL} + N_{RQ})$, where $N_{RL}$ ($N_{RQ}$) is the number of photons attributable to RL (RQ) QSOs. Since the non-VLA subsamples contribute only detections, we must calculate $R_{RL}$ for the VLA-observed subsample and assume that it applies to the overall LBQS/RASS. Where the largest change in $\langle \Gamma \rangle$ occurs, between QSO redshift classes of $\langle z \rangle = 0.8$ and 1.7, $R_{RL}$ decreases from 27.3% to 20.9%. Such a decrease would be expected to produce an *increase* in $\langle \Gamma \rangle$, opposite to the trend observed. Thus, if this fraction for the VLA subsample is representative of the fraction for the whole LBQS/RASS sample in these redshift bins, the apparent flattening (hardening) of the X-ray spectra is not due to an increasing contribution of RL QSOs.

2) Absorption over and above that due to our Galaxy. The small number of total counts necessitates our assumption here that $N_H^{tot} \approx N_H^{Gal}$, leaving two fit parameters $\Gamma$, and $I$. This assumption is consistent with results from most previous X-ray studies of low redshift QSOs when fitting a single power law model. Exceptions occur in some sources with "warm absorbers" (e.g., Nandra & Pounds 1992, Laor et al. 1994), and in a few intermediate to high-redshift RL QSOs (e.g., Elvis et al. 1994a,b). Consistent with these latter results, it is possible that the observed trend in $\langle \Gamma \rangle$ is due to absorption along the line of sight that increases toward higher redshift. Such absorption would reduce the observed soft X-ray emission, whereby our one-parameter fits here would yield a harder apparent photon index. If absorption is the sole cause of the observed trends, then it appears to become significant only for redshifts greater than about one. This would imply a strong evolution of the intrinsic absorption properties of QSOs with look-back time, or (due to the strong correlation between redshift and luminosity in the LBQS) with luminosity. Since strong X-ray absorption in high luminosity QSOs is unusual (e.g., Wilkes et al. 1987, Williams et al. 1992), intervening absorbers are more likely, as argued by Elvis et al. (1994a). The latter study was of RL QSOs only. Assuming that RL QSOs are probably giant ellipticals in clusters, those authors postulate that the absorption is associated with cold intracluster gas (e.g., cooling flows). Since our overall counts are dominated by RQ QSOs, and since there is some weak additional evidence for a decrease in photon index with redshift for the RQ class alone, our work suggests that this apparent hardening is not solely due to associated absorption in the presumed cluster environments of RL QSOs. A hypothesis not inconsistent with our results is that of intervening absorbers such as damped Ly$\alpha$ clouds at $0.3 < z < 1.9$, (Elvis et al. 1994a, Vikhlinin et al. 1995).

3) A soft excess due to a double power law. In many QSOs, at least two emission components are discernible in the X-rays: a flat nonthermal power law ($\Gamma \approx 1.7 - 2.0$; e.g., Turner & Pounds 1989), and a steeper soft X-ray component ($\Gamma \approx 2.8$; Fiore et al. 1993), with the spectral break between these two components somewhere between 0.3 and 1 keV (e.g., Jackson, Browne, & Warwick 1993). A break energy in this range would produce a hardening of the RASS spectra in the redshift range $1 < z < 2$ as observed, since the soft component would shift out of the *ROSAT* band between those redshifts.

4) The apparent hardening with redshift of the best-fit power law photon index $\langle \Gamma \rangle$ for QSOs in the LBQS/RASS sample might also be evidence for the evolution of the physical conditions in the QSO nucleus. For instance, the models of Haardt & Maraschi (1993) predict that larger black hole masses yield *softer* X-ray spectra. This would be consistent with scenarios in which the black hole mass grows with time by accretion in long-lived QSOs, as discussed in Bechtold et al. (1994).

In summary, we have studied here the soft X-ray spectral properties of QSOs in a large, homogeneous sample, the LBQS/RASS. We find that the QSOs in our sample with strong optical FeII emission have a mean X-ray spectral photon index $\langle \Gamma \rangle$ consistent with that of a control sample. We find some evidence for a difference in $\langle \Gamma \rangle$ between radio-loud and radio-quiet QSOs, and between QSOs at low and high redshift. Due to the moderate PSPC spectral resolution, and the relatively short exposure times of the RASS, the data here are not adequate to distinguish between the variety of proposed causes of the observed decrease in $\langle \Gamma \rangle$ with redshift. We have, however, demonstrated the utility of a novel spectral stacking technique which may improve this situation by facilitating rigorous analysis of co-added X-ray spectral data from samples of QSOs with deep pointed X-ray observations, even across wide regions of they sky.

## 7 ACKNOWLEDGEMENTS

The *ROSAT* project is supported by the Bundesministerium für Forschung und Technologie (BMFT). We thank our colleagues from the *ROSAT* group for their support. NS acknowledges a Max Planck Fellowship. This research was supported through NASA Grant NAG5-1623. Support for PJG was provided by the NSF through grant INT 9201412, and by NASA through Grant HF-1032.01-92A awarded by the Space Telescope Science Institute, which is operated by the Association of Universities for Research in Astronomy, Inc., under NASA contract NAS5-26555. The LBQS is supported by NSF grant 93-20715 to CBF. Thanks also to Bernhard Beck for assistance in preparation of the manuscript.

## APPENDIX A: FITTING THE MODEL TO THE DATA

To determine the best-fit slope for a power law model, whether for an individual source spectrum or a stacked spectrum, we fold the model through the PSPC response function for comparison to the data. If, after stacking, the total counts are still too low for a full least-squares fit of a power law model to the observed X-ray spectrum, we can use the correspondence between hardness ratios and X-ray photon index to derive estimates for the latter quantity. We use two *adaptive* hardness ratios (AHRs):

$HR1 = (B - A)/(A + B)$

and

$HR2 = (C - D)/(C + D)$ .

$A, B, C$, and $D$ are the estimated number of (vignetting-corrected) source counts in the following channels: $A : 8 \to b$, $B : (b+1) \to 235$, $C : (b+1) \to c$, and $D : (c+1) \to 235$. Here the energy of a given channel in keV is approximately (channel no.)/100. The channel numbers $b$ and $c$ are adapted for each spectrum to yield approximately equal S/N ratios for $A, C$, and $D$. This method allows calculation of a hardness ratio even when the source has an extremely steep or heavily absorbed spectrum. In practice, we use hardness ratios only when S/N$>$ 4 in three bins ($A, C$, and $D$). Least-squares fits are used when the S/N ratio exceeds 5 in at least 6 bins. To determine the photon index $\Gamma$ of the assumed power law from hardness ratios $HR1$ and $HR2$, we fold through the PSPC response function a power law model with fixed Galactic absorption and derive the expected hardness ratios. We then vary the input $\Gamma$ to minimize the sum of the quadratic differences between the measured and calculated hardness ratios, and quote errors corresponding to 68% confidence levels for a single parameter. A detailed description with tests of the method will be published in Schartel, Walter, & Fink (1995). More detailed tests can be found in Schartel (1994)

**FIGURE CAPTIONS**

Figure 1. Photon count rate as a function of energy. As an example of our spectral stacking technique, we show the count rate for the total stacked spectrum (source plus background) of the 72 detected QSOs in the LBQS/RASS with $0.2 < z < 1$. Vertical bars represent the errors in count rate, while horizontal bars show the energy range of the photons stacked in each energy bin. The area-normalized, stacked background spectrum for this sample is also shown. Poisson count rate errors for the stacked background are of similar size to the line thickness.

Figure 2. Fits using 4 models to stacked X-ray spectra of 3 simulated QSO samples. EWA is best adapted to samples including non-detections, observed through different absorbing columns. The fractional difference spectra, (fit − input)/input, are shown in lower panels. a) Input spectra have a single power law slope ($\Gamma = -2.5$) but a range of $N_H$. EWA excels here, since the EWA fit is essentially identical to the input spectrum. b) Input spectra all have $N_H = 2.5 \times 10^{20}$ cm$^{-2}$ but span a range of $\Gamma$. Fits using $\langle N_H \rangle^{EW}$ and $\langle N_H \rangle^{CW}$ are identical to the input $N_H$ in this case, and are omitted from the plot. The EWA model yields a superior fit to a model with free $N_H$. c) Input spectra have a range of both $\Gamma$ and $N_H$. EWA is marginally superior to the model with counts-weighted $\langle N_H \rangle^{CW}$. However, the latter model can only be used when all sources are detected.

Figure 3. X-ray photon index as a function of redshift. The best-fit (0.1 - 2.4 keV) mean power law photon index $\langle \Gamma \rangle$ from stacking is shown as a function of redshift for LBQS QSOs in 4 redshift ranges as observed in the RASS (filled circles with solid bars). The mean photon index for all 908 LBQS QSOs is also shown (filled circle with dashed bars). Vertical bars are the uncertainty in the photon index, while horizontal bars show the redshift range of the QSOs measured.